# Research on Flexibility Margin of Electric-Hydrogen Coupling Energy Block Based on Model Predictive Control

**Zijiao Han**[1,2], **Shun Yuan**[1,3], **Yannan Dong**[1], **Shaohua Ma**[1*], **Yudong Bian**[4], **Xinyu Mao**[5]

[1] Institute of Electrical Engineering, Shenyang University of Technology, Shenyang 110870, China.

[2] Liaoning Electric Power Company, State Grid Corporation of China, Shenyang 110004, China.

[3] National Energy Administration, Beijing 100085, China.

[4] Jiangmen Power Supply Bureau of Guangdong Power Grid Corporation, Jiangmen 529000, China.

[5] School of Electrical Engineering, Northeast Electric Power University, Jilin 132012, China.

**\* Correspondence:** Shaohua Ma (mash_dq@sut.edu.cn)

**Abstract:** Hydrogen energy plays an important role in the transformation of low-carbon energy, and electric hydrogen coupling will become a typical energy scenario. Aiming at the operation flexibility of low-carbon electricity hydrogen coupling system with high proportion of wind power and photovoltaic, this paper studies the flexibility margin of electricity hydrogen coupling energy block based on model predictive control (MPC). By analyzing the power exchange characteristics of heterogeneous energy, the homogenization models of various heterogeneous energy sources are established. According to the analysis of power system flexibility margin, three dimensions of flexibility margin evaluation indexes are defined from the dimension of system operation, and an electricity hydrogen coupling energy block scheduling model is established. The model predictive control algorithm is used to optimize the power balance operation of the electro hydrogen coupling energy block, and the flexibility margin of the energy block is quantitatively analyzed and calculated. Through the example analysis, it is verified that the calculation method proposed in this paper can not only realize the on-line power balance optimization of electric hydrogen coupling energy block, but also effectively quantify the operation flexibility margin of electric hydrogen coupling energy block.

## 1 Introduction

As the penetration ratio of renewable energy increases year by year, there is a reverse change between supply and demand of power system flexibility, and the phenomenon of renewable energy power limitation is increasing day by day [1, 2]. This transition undoubtedly brings huge economic and environmental benefits, but also bring huge challenges to the secure and stable operation of the today's power system due to the uncertainties of renewable power generations [3-5]. As a clean secondary energy [6], it is particularly important for hydrogen to fully call all links of source charge storage to participate in flexibility adjustment and balance through electric hydrogen coupling power generation system and cooperate with the consumption of renewable energy [7, 8]. Power to gas (P2G), as links of multi-energy carriers, has been successfully adopted to strengthen the coupling of different energy systems in [9]. Hydrogen-electric coupling will greatly contribute to promote renewable energy consumption and build a low-carbon sustainable modern energy system [10, 11]. In order to solve the problems of large differences in heterogeneous energy models and difficult flexibility margin analysis

**Research on Flexible Margin of Electric-Hydrogen Coupling Energy Block Based on MPC**

in electro hydrogen coupling module, it is urgent to build an effective electric hydrogen coupling contract qualitative energy module and carry out flexibility margin calculation.

At present, many experts and scholars around the world have carried out research on the modeling and flexibility margin analysis of electro hydrogen coupling module. Literature [12] established a linear model of wind hydrogen hybrid energy system based on empirical and physical relations, and analyzed the performance of hybrid energy system through digital simulation. Reference [13] carried out conceptual modeling of hybrid renewable energy system based on wind energy / electrolyzer / proton exchange membrane fuel cell. At the same time, based on the thermodynamic, electrochemical and mechanical models of different components of the hybrid system, the energy analysis framework of the hybrid system is established. And the model is used for simulation analysis to calculate the influence of different operating parameters on energy application efficiency. Reference [14] established a renewable energy hybrid system model composed of photovoltaic cells, wind turbines, battery energy storage systems and diesel power generation systems. And three cases are set to analyze the impact of system performance and intermittent renewable energy on the system. In addition, in the operation and analysis of multi-energy power system. Research institutions at home and abroad have carried out relevant research on the flexibility evaluation of two or a few energy resource systems. Starting from the dimension of flexibility planning, document [15] defines power system flexibility as the ability of the system to allocate its resources when the system network or load changes, and takes this flexibility as one of the indicators to evaluate whether the planning scheme is reasonable. The flexibility of the power system described in the literature [16] is the ability to quickly adapt and restore stability when the system undergoes unpredictable fluctuations or changes under certain economic constraints. Document [17] qualitatively describes the system flexibility in this area by using the switching capacity of hydropower and thermal power generation, and the economic benefits brought by hydro thermal power switching can measure the advantages and disadvantages of switching schemes under different operating states. References [18, 19] define the unit flexibility index and system flexibility index according to the unit flexibility parameters, and establishes the unit combination model considering unit expansion with the goal of minimizing the total cost.

The above-mentioned literature lacks homogeneous equivalent modeling of heterogeneous energy systems and multi-dimensional quantitative understanding of flexibility. Moreover, the flexibility evaluation index is one-sided and single, and it is difficult for the flexibility evaluation to reveal the flexibility boundary of the system operation. Therefore, this paper analyzes the power characteristics of the heterogeneous energy of electric and hydrogen, and carries out the homogeneity modeling of the heterogeneous energy. And from the three dimensions of climbing, power and energy, the energy block balance criterion and flexibility margin index of electric-hydrogen coupling are proposed. The flexibility margin of the electric-hydrogen coupling energy block is analyzed, and the power balance optimization calculation method of the energy block based on model predictive control (MPC) is proposed. Finally, combined with the historical data of a new energy demonstration area in my country, the calculation method of the flexibility margin of the electricity-hydrogen coupling energy block proposed in this paper was verified.

## 2    Electro-hyderogen coupling energy block and homogenization modeling

Based on the existing gas turbine, wind power and photovoltaic structures, combined with the characteristics of electrolytic cell and fuel cell, an electric hydrogen coupling energy block is constructed, as shown in Figure 1.

It can be seen from Figure 1 that in the energy block composed of a variety of heterogeneous energy sources, wind power, photovoltaic and gas units are coupled with hydrogen energy storage devices. At the same time, the balance supply of electric load and hydrogen load is realized through internal energy regulation. The mutual conversion between hydrogen energy and electric energy is the

# Research on Flexible Margin of Electric-Hydrogen Coupling Energy Block Based on MPC

medium connecting various heterogeneous energy models in the electric hydrogen coupling energy block. For the outside of the electric hydrogen coupling energy block, the energy module is supplied by wind energy, solar energy, gas and hydrogen, and the supply and demand balance of source load is met through internal energy conversion. For the inside of the electric hydrogen coupling energy block, wind airports and photovoltaic stations convert the supplied energy into electric energy. When the power output is higher than the electric load demand, the electrolytic cell absorbs the excess power of the system, converts the electric energy into hydrogen energy and stores it. When the power output is less than the electric load demand, the fuel cell will supplement the missing power. The hydrogen storage capacity is determined by the electric load and hydrogen load. The battery energy storage can balance the climbing power and make up for the power shortage.

The electro hydrogen coupling energy block involves a variety of heterogeneous energy units with different physical media and technical characteristics. For its complex parameters and boundary conditions, it is urgent to establish a unified model for analysis.

Combined with the structure of electro hydrogen coupling energy block and considering the power exchange characteristics of heterogeneous energy, the model framework is shown in Figure 2.

As can be seen from the figure, the supply and demand side of the model includes energy supply process ($\zeta_i > 0$), demand process ($\zeta_i < 0$), and spillover energy ($w_i \geq 0$). The model system side includes the homogenization model and the conversion and transmission process of electric energy and hydrogen energy. By changing the model parameters, the model can characterize the elements in the energy block.

The model set $i \in \{w, pv, b, h, f, \cdots\}$, where $w$ represents wind farm, $pv$ represents photovoltaic station, $b$ represents battery energy storage, $h$ represents hydrogen energy storage system and $f$ represents gas unit. The general formula of model $i$ is:

$$C_i dx_i = \eta_{ex,i} \xi_i - \eta_{gen,i} p_{gen,i} \Delta t + \eta_{load,i} p_{load,i} \Delta t - w_i \tag{1}$$

Where, $C_i$ is the energy storage capacity, $dx_i$ is the change in state of charge (SOC) of energy storage, $\eta_{gen,i}$ is the output efficiency, $p_{gen,i}$ is the output power, $\eta_{load}$ is the load efficiency, $p_{load,i}$ is the load power, $\eta_{ex,i}$ is the conversion efficiency of external energy, $\zeta_i$ is the output / consumption of external energy, and $w_i$ is the overflow energy, $\Delta t$ is the dispatching cycle of energy storage.

According to the existence of energy storage, the heterogeneous energy models in the energy module are divided into two categories. The renewable energy power generation model without energy storage ($C_i dx_i = 0$) is as follows:

The wind farm model is formulated by

$$C_w dx_w = 0 = \xi_w - \eta_{gen,w} p_{gen,w} \Delta t - w_w \tag{2}$$

Where, $\zeta_w$ is the wind energy collected by the wind farm for the collection and calculation of wind speed data, $\eta_{gen,w}$ is the power generation efficiency of the wind farm itself, $p_{gen,w}$ is the output power of the wind farm, and $w_w$ is the waste air volume in the dispatching cycle.

The PV station model is given as follows:

$$C_{pv} dx_{pv} = 0 = \xi_{pv} - \eta_{gen,pv} p_{gen,pv} \Delta t - w_{pv} \tag{3}$$

Where, $\zeta_{pv}$ is the solar energy collected by the photovoltaic station for the collection and calculation of irradiation data, $\eta_{gen,pv}$ is the power generation efficiency of the photovoltaic power plant itself, $p_{gen,pv}$ is the output power of the photovoltaic station, and $w_{pv}$ is the light rejection in the dispatching cycle.

# Research on Flexible Margin of Electric-Hydrogen Coupling Energy Block Based on MPC

The heterogeneous energy model with energy storage ($C_i dx_i \neq 0$) can be depicted in the following section.

The battery energy storage model is:

$$C_b dx_b = \eta_{load,b} p_{load,b} \Delta t - \eta_{gen,b} p_{gen,b} \Delta t \tag{4}$$

Where, $C_b$ represents the energy storage capacity of the battery, $\eta_{load,b}$ represents the charging efficiency of the battery, $p_{gen,b}$ represents the charging power of the battery, $\eta_{gen,b}$ represents the discharge efficiency of the battery, and $p_{gen,b}$ represents the discharge power of the battery.

Hydrogen energy storage system model:

$$C_h dx_h = \xi_h + \eta_{load,h} p_{load,h} \Delta t - \eta_{gen,h} p_{gen,h} \Delta t \tag{5}$$

Where, $C_h$ represents the capacity of the hydrogen storage tank, $\zeta_h$ represents the hydrogen supply and demand, when $\zeta_h>0$, it represents the hydrogen supply, when $\zeta_h<0$, it represents the hydrogen demand, $\eta_{load,h}$ represents the hydrogen production efficiency of the electrolytic cell, $p_{load,h}$ represents the power of the electrolytic cell, $\eta_{gen,h}$ represents the power generation efficiency of the fuel cell, and $p_{gen,h}$ represents the output power of the fuel cell.

The gas unit model is given as follows:

$$C_f dx_f = \xi_f - \eta_{gen,f} p_{gen,f} \Delta t \tag{6}$$

Where, $C_f$ represents the capacity of the gas storage pipe, $\zeta_f$ is the gas input, $\eta_{gen,f}$ is the gas power generation efficiency, $p_{gen,f}$ is the output power of the gas unit.

The constraints to be met by the model are as follows:

1) Climbing constraint: unit and load climbing rate constraint

$$dp_{load,i,\min} \leq dp_{load,i} \leq dp_{load,i,\max} \tag{7}$$

Where, $dp_{load,i,\min}$ and $dp_{load,i,\max}$ are the lower limit and upper limit of load climbing rate, respectively.

$$dp_{gen,i,\min} \leq dp_{gen,i} \leq dp_{gen,i,\max} \tag{8}$$

Where, $dp_{gen,i,\min}$ and $dp_{gen,i,\max}$ are the lower limit and upper limit of unit climbing rate, respectively.

2) Output power constraints: system load and output constraints:

$$0 \leq kp_{load,i,\min} \leq kp_{load,i} \leq kp_{load,i,\max} \tag{9}$$

Where, $p_{load,i,\min}$ and $p_{load,i,\max}$ are respectively the upper and lower limit values of load demand, and each efficiency parameter is a variable, which changes with the change of input/output power, where $\eta_{gen,i}=f(p_{gen,i})$, $\eta_{load,i}=f(p_{load,i})$, $\eta_{ex,i}=f(\zeta_i)$, $k$ is a binary variable. Here, $k=1$ or $0$ represents load is running or not running, respectively.

$$0 \leq Xp_{gen,i,\min} \leq Xp_{gen,i} \leq Xp_{gen,i,\max} \tag{10}$$

Where, $p_{gen,i,\min}$ and $p_{gen,i,\max}$ are respectively the upper limit and lower limit of unit output, and each efficiency parameter is a variable, which changes with the change of input/output power, where $\eta_{gen,i}=f(p_{gen,i})$, $\eta_{load,i}=f(p_{load,i})$, $\eta_{ex,i}=f(\zeta_i)$, $X$ is a binary variable. Here, $X=1$ or $0$ represents a unit is running or not running, respectively.

3) Energy storage constraints: state of charge and energy storage capacity constraints:

$$\begin{cases} C_i > 0, & x_{\min} \leq x_i \leq x_{\max} \\ C_i = 0 \end{cases} \quad (11)$$

Where $x_{\min}$ and $x_{\max}$ are the upper and lower limits of the state of charge respectively. When there is energy storage, $C > 0$, otherwise, when $C = 0$, there is no state of charge constraint.

4) Renewable energy abandonment constraints: energy input/consumption and spillover energy constraints:

$$\begin{cases} \zeta_i > 0, w_i \geq 0, \zeta_i - w_i \geq 0 \\ \zeta_i < 0 \end{cases} \quad (12)$$

Where $\zeta_i > 0$ represents the external input of energy, $w_i \geq 0$ represents the existence of overflow energy (wind and light abandonment), and $\zeta_i < 0$ represents the external consumption of energy.

## 3 Flexbility margin analysis and evaluation index of energy block

### 3.1 Energy block flexibility margin analysis

The flexibility demand of energy block mainly comes from the intermittence and fluctuation of load and renewable energy, and has certain directionality. Flexibility can be divided into upward flexibility requirements and downward flexibility requirements. When the system has upward flexibility demand ($p_{net} \leq 0$), the flexibility resource response is gas unit, battery energy storage and discharge, fuel cell and load rejection. When the system has downward flexibility demand ($p_{net} \geq 0$), the flexibility resource response is battery energy storage charging, electrolytic cell and renewable energy abandonment.

Aiming at the balance between supply and demand of flexibility in electro hydrogen coupling energy block. This paper will analyze flexibility from three dimensions: climbing power, output power and power supply [20-22].

As shown in Figure 3, four operation points are set to analyze the flexibility margin of the system at $t_2$ time. Operation points A and B have downward flexibility requirements, and the flexibility provided at $t_1$ time is $r_p^-$, $p_p^-$, $e_p$, while the downward flexibility required by operation point a is $r_n^-$, $p_n^-$, $e_n^-$. It can be seen from the figure that the flexibility margin required for operation point a is outside the system flexibility margin envelope, $r_p^-$, $p_p^-$, $e_p^- < r_n^-$, $p_n^-$, $e_n^-$ and the system can't meet the flexibility requirements of operation point A. The flexibility margin required by operation point B is within the system flexibility margin envelope, and the system can meet the flexibility margin requirements of operation point B.

Operation points C and D have upward flexibility requirements. The operating point C is within the flexibility margin envelope, which can meet the requirements of flexibility margin. Operating point D is outside the envelope of flexibility margin, $r_p^+$, $p_p^+$, $e_p^+ > r_n^+$, $p_n^+$, $e_n^+$, which can't meet the requirements of flexibility margin.

The flexibility balance criterion of electric hydrogen coupling energy block is:

**Research on Flexible Margin of Electric-Hydrogen Coupling Energy Block Based on MPC**

$$\begin{cases} \sum_i r^+_{p,i,t} \geq r^+_{n,t}, \sum_i r^-_{p,i,t} \leq r^-_{n,t} \\ \sum_i p^+_{p,i,t} \geq p^+_{n,t}, \sum_i p^-_{p,i,t} \leq p^-_{n,t} \\ \sum_i e^+_{p,i,t} \geq e^+_{n,t}, \sum_i e^-_{p,i,t} \leq e^-_{n,t} \end{cases} \quad (13)$$

Where, $r^+_{p,i,t}$, $r^-_{p,i,t}$ are the uphill power and downhill power provided by unit $i$ at time $t$, $r_{n,t}$ is the uphill power required by the system at time $t$, $p^+_{p,i,t}$ and $p^-_{p,i,t}$ are the up output power and down output power provided by unit $i$ at time $t$, $p_{n,t}$ is the output power required by the system at time $t$, $e^+_{p,i,t}$, $e^-_{p,i,t}$ is the upper power supply and lower power supply provided by unit $i$ at time $t$, and $e_{n,t}$ is the power supply required by the system at time $t$.

### 3.2 Energy block evaluation index

Calculate the flexibility margin required at each running time and the upper and lower limits of the flexibility margin provided by the energy block, draw the flexibility margin envelope, and analyze the boundary of the system flexibility margin [23, 24]. And through the analysis of the flexibility margin required by the operating point, the expectation of insufficient flexibility margin in each dimension is defined as an index to measure the flexibility margin of the energy block.

To enable it to effectively reflect the system's ability to respond to changes in net load, the indicators are as follows:

1) The climbing power does not meet the index $E_{IR}$, which indicates the expected value of the difference between the up / down climbing power provided by the system and the actual demand climbing power within the operation day.

$$E_{IR} = \rho_s (\sum_{t=1}^{N_L} \left| dP_{net,t} - \sum_{i=1}^{N_G} dP_{gen,i} \right| + \left| \sum_{i=1}^{N_L} dP_{load,i} - dP_{net,t} \right|) \quad (14)$$

In formula (14):

$$\rho_s = \frac{\beta_{s,t}}{N_T} \quad (15)$$

Where, $dP_{net,t}$ represents the net load climbing rate at time $t$, $\rho_s$ represents the probability of insufficient flexibility margin in scenario $s$, $N_T$ represents the number of system dispatching intervals, $\beta_{s,t}$ represents the number of insufficient flexibility margin, $N_G$ represents the number of generator units, and $N_L$ represents the number of loads.

2) The output power does not meet the index $E_{IO}$, which indicates the expected value of the difference between the up / down power provided by the system and the actual demand power within the operation day.

$$E_{IO} = \rho_s (\sum_{t=1}^{N_T} \left| P_{net,t} - \sum_{i=1}^{N_G} P_{gen,i} \right| + \left| \sum_{i=1}^{N_L} P_{load,i} - P_{net,t} \right|) \quad (16)$$

Where, $P_{net,t}$ represents the net load power at time $t$.

# Research on Flexible Margin of Electric-Hydrogen Coupling Energy Block Based on MPC

3) The provided electric energy does not meet the $E_{IC}$ index, which indicates the expected value of the difference between the up / down regulated electric energy provided by the system energy storage unit and the actual demand electric energy within the operation day.

$$E_{IC} = \rho_s (\sum_{t=1}^{N_T} \left| \int_t^{t+1} P_{net,t} dt - \sum_{i=1}^{N_G} \int_t^{t+1} P_{gen,i} dt \right| + \left| \sum_{i=1}^{N_L} \int_t^{t+1} P_{load,i} dt - \int_t^{t+1} P_{net,t} dt \right|) \tag{17}$$

The above three indicators show the expectation of insufficient flexibility margin as that the flexibility margin unit can't meet the change of system static load.

## 4  Online solution of energy block flexibility margin based on MPC

The state space of linear discrete system is expressed as:

$$\begin{cases} x(k+1) = Ax(k) + Bu(k) + Dd(k) \\ y(k) = Cx(k) \end{cases} \tag{18}$$

Where, $x$ is the state variable, $u$ is the control variable, $d$ is the disturbance variable, $y$ is the controlled output, $A$, $B$ and $C$ are the system matrix, control matrix and disturbance matrix respectively, $C$ is the output matrix, $k$ is the current time, and $k+1$ is the next time.

Bring equations (2)-(6) into equation (18), and the specific form of state space expression is:

$$\begin{cases} x_w(k+1) = \zeta_w(k) - \eta_{gen,w} P_{gen,w}(k)\Delta t - w_w(k) \\ x_{pv}(k+1) = \zeta_{pv}(k) - \eta_{gen,pv} P_{gen,pv}(k)\Delta t - w_{pv}(k) \\ x_b(k+1) = x_b(k) + \dfrac{1}{C_b}(\eta_{load,b} P_{load,b}(k)\Delta t \\ \qquad\qquad - \eta_{gen,b} P_{gen,b}(k)\Delta t) \\ x_h(k+1) = x_h(k) + \dfrac{1}{C_h}(\zeta_h(k) + \eta_{load,h} P_{load,h}(k)\Delta t \\ \qquad\qquad - \eta_{gen,h} P_{gen,h}(k)\Delta t) \\ x_f(k+1) = x_f(k) + \zeta_f(k) - \eta_{gen,f} P_{gen,f}(k) \end{cases} \tag{19}$$

The vector formed by the energy storage SOC is selected as the state variable, namely:

$$x(k) = \left[ x_w(k), x_{pv}(k), x_e(k), x_h(k), x_f(k) \right]^T \tag{20}$$

The vector composed of output, load power and overflow energy of each unit is taken as the control variable, namely:

$$u(k) = \left[ p_{gen,w}(k), p_{gen,pv}(k), p_{gen,b}(k), p_{load,b}(k), \\ p_{gen,h}(k), p_{load,h}(k), p_{gen,f}(k), w_w(k), w_{pv}(k) \right]^T \tag{21}$$

# Research on Flexible Margin of Electric-Hydrogen Coupling Energy Block Based on MPC

Take wind power, photovoltaic, hydrogen, and coal energy supply vectors as disturbance variables, namely:

$$d(k) = [\xi_w(k), \xi_{pv}(k), \xi_h(k), \xi_f(k)]^T \tag{22}$$

Take the vector composed of battery energy storage and hydrogen energy storage SOC as the output variable, namely:

$$y(k) = [x_b(k), x_h(k)]^T \tag{23}$$

The system matrix, control matrix, output matrix and disturbance matrix of the energy module are respectively:

$$A = \begin{bmatrix} 0 & 0 & 0 & 0 & 0 \\ 0 & 0 & 0 & 0 & 0 \\ 0 & 0 & 1 & 0 & 0 \\ 0 & 0 & 0 & 1 & 0 \\ 0 & 0 & 0 & 0 & 1 \end{bmatrix} \tag{24}$$

$$B^T = \begin{bmatrix} -\eta_{gen,w} & 0 & 0 & 0 & 0 \\ 0 & -\eta_{gen,pv} & 0 & 0 & 0 \\ 0 & 0 & \dfrac{-\eta_{gen,b}\Delta t}{C_b} & 0 & 0 \\ 0 & 0 & \dfrac{\eta_{load,e}\Delta t}{C_e} & 0 & 0 \\ 0 & 0 & 0 & \dfrac{-\eta_{gen,h}\Delta t}{C_h} & 0 \\ 0 & 0 & 0 & \dfrac{\eta_{load,h}\Delta t}{C_h} & 0 \\ 0 & 0 & 0 & 0 & -\eta_{gen,f} \\ 1 & 0 & 0 & 0 & 0 \\ 0 & 1 & 0 & 0 & 0 \end{bmatrix} \tag{25}$$

$$C = \begin{bmatrix} 0 & 0 & 1 & 0 & 0 \\ 0 & 0 & 0 & 1 & 0 \end{bmatrix} \tag{26}$$

$$D = \begin{bmatrix} 1 & 0 & 0 & 0 \\ 0 & 1 & 0 & 0 \\ 0 & 0 & 0 & 0 \\ 0 & 0 & 1 & 0 \\ 0 & 0 & 0 & 1 \end{bmatrix} \tag{27}$$

# Research on Flexible Margin of Electric-Hydrogen Coupling Energy Block Based on MPC

The error between the estimated output value of spilled energy and stored energy SOC and the daily planned value is the smallest, and the control active power regulation increment of each unit is the smallest. The objective function of the electro hydrogen coupling energy block is:

$$\min J(k) = \sum_{j=1}^{N_p} \| y(k+j) - y_{ref}(k+j) \|_Q^2 + \sum_{j=0}^{N_c} \| \Delta u(k+j) \|_R^2 \tag{28}$$

Where $y_{ref}$ is the reference trajectory of state quantity, $Q$ and $R$ are the error and input weighting matrix respectively, and $N_p$ and $N_c$ are respectively the prediction time domain and control time domain.

The constraints are as follows:

$$\text{S.t.} \quad \sum_{i=1}^{N_G} p_{gen,i} - \sum_{i=1}^{N_L} p_{load,i} - P_{load}^t = 0 \tag{29}$$

$$\Delta u_{min} \leq \Delta u \leq \Delta u_{max} \tag{30}$$

$$u_{min} \leq u \leq u_{max} \tag{31}$$

$$x_{min} \leq x \leq x_{max} \tag{32}$$

Where $P^t_{load}$ represents the system load demand power at time $t$.

Define vector $E(k)$ as the deviation between the system free response and the future target trajectory:

$$E(k) = Y_{ref}(k) - M_{x_1} x(k) - M_{u_1} u(k-1) \tag{33}$$

In formula (33):

$$M_{x_1} = \begin{bmatrix} CA \\ CA^2 \\ \vdots \\ CA^{N_c} \\ CA^{N_c+1} \\ \vdots \\ CA^{N_p} \end{bmatrix} \quad M_{u_1} = \begin{bmatrix} CB \\ CAB + CB \\ \vdots \\ \sum_{i=0}^{N_c-1} CA^i B \\ \sum_{i=0}^{N_c} CA^i B \\ \vdots \\ \sum_{i=0}^{N_p-1} CA^i B \end{bmatrix} \tag{34}$$

Where, see equations (24) to (26) for matrices $A$, $B$ and $C$ substitute equation (33) into equation (28) to obtain:

# Research on Flexible Margin of Electric-Hydrogen Coupling Energy Block Based on MPC

$$\begin{aligned} J_k &= \|Y(k) - Y_{ref}(k)\|_Q^2 + \|\Delta U(k)\|_R^2 \\ &= \|M_{\Delta u_1} \Delta U(k) - E(k)\|_Q^2 + \|\Delta U(k)\|_R^2 \\ &= \Delta U^T(k) \left[ M_{\Delta u_1}^T Q M_{\Delta u_1} + R \right] \Delta U(k) \\ &\quad - 2E^T(k) Q M_{\Delta u_1} \Delta U(k) + E^T(k) Q E(k) \end{aligned} \quad (35)$$

The above formula is written as the standard form of secondary planning:

$$J_k = \frac{1}{2} \Delta U^T(k) H(k) \Delta U(k) + f^T \Delta U(k) \quad (36)$$

In formula (36):

$$\begin{aligned} H &= 2\left( M_{\Delta u_1}^T Q M_{\Delta u_1} + R \right) \\ f &= -2\left( M_{\Delta u_1}^T Q E(k) \right) \end{aligned} \quad (37)$$

$$M_{\Delta u_1} = \begin{bmatrix} CB & \cdots & 0 \\ CAB + CB & \cdots & 0 \\ \vdots & \ddots & \vdots \\ \sum_{i=0}^{N_c-1} CA^i B & \cdots & B \\ \sum_{i=0}^{N_c} CA^i B & \cdots & CAB + CB \\ \vdots & \vdots & \vdots \\ \sum_{i=0}^{N_p-1} CA^i B & \cdots & \sum_{i=0}^{N_p-N_c} CA^i B \end{bmatrix} \quad (38)$$

The constraint condition of equation (36) is:

$$\tilde{A} \Delta U(k) \leq \tilde{b} \quad (39)$$

In formula (39):

$$\tilde{A} = \begin{bmatrix} \Pi & -\Pi & \Lambda & -\Lambda & M_{\Delta u_2} & -M_{\Delta u_2} \end{bmatrix}^T \quad (40)$$

$$\tilde{b} = \begin{bmatrix} \Delta U_{max} \\ -\Delta U_{min} \\ U_{max} - \Psi u(k-1) \\ -U_{min} + \Psi u(k-1) \\ Y_{max} - M_{x_2} x(k) - M_{u_2} u(k-1) \\ -Y_{min} + M_{x_2} x(k) + M_{u_2} u(k-1) \end{bmatrix} \quad (41)$$

# Research on Flexible Margin of Electric-Hydrogen Coupling Energy Block Based on MPC

$$\Lambda = \begin{bmatrix} I & 0 & \cdots & 0 \\ I & I & \cdots & 0 \\ \vdots & \vdots & \ddots & \vdots \\ I & I & \cdots & I \end{bmatrix}_{N_c \times N_c} \quad \Psi = \begin{bmatrix} I \\ I \\ \vdots \\ I \end{bmatrix}_{N_c \times 1} \tag{42}$$

$$M_{x_2} = \begin{bmatrix} CA \\ CA^2 \\ \vdots \\ CA^{N_c} \\ CA^{N_c+1} \\ \vdots \\ CA^{N_p} \end{bmatrix}, M_{u_2} = \begin{bmatrix} CB \\ CAB + CB \\ \vdots \\ \sum_{i=0}^{N_c-1} CA^i B \\ \sum_{i=0}^{N_c} CA^i B \\ \vdots \\ \sum_{i=0}^{N_p-1} CA^i B \end{bmatrix} \tag{43}$$

$$M_{\Delta u_2} = \begin{bmatrix} CB & \cdots & 0 \\ CAB + CB & \cdots & 0 \\ \vdots & \ddots & \vdots \\ \sum_{i=0}^{N_c-1} CA^i B & \cdots & B \\ \sum_{i=0}^{N_c} CA^i B & \cdots & CAB + CB \\ \vdots & \vdots & \vdots \\ \sum_{i=0}^{N_p-1} CA^i B & \cdots & \sum_{i=0}^{N_p-N_c} CA^i B \end{bmatrix} \tag{44}$$

Where, $I$ is the identity matrix of $m$ dimension, and $m$ is the number of control quantities.

According to the above derivation process, it can be transformed into a quadratic programming problem for solution. As shown in formula (45), the first item of the solved control sequence is applied to the system, and the control quantity is executed until the next time. At the new time, the system re predicts the output of the next time domain according to the state information. Then the next new control increment sequence is obtained through the optimization process. This cycle is repeated until the system completes the control process.

$$\begin{cases} \Delta u(k) = \begin{bmatrix} I & 0 & \cdots & 0 \end{bmatrix} \Delta U(k) \\ u(k) = u(k-1) + \Delta u(k) \end{cases} \tag{45}$$

Where $k$-1 is the last time.

Draw the flexibility margin envelope according to the calculated control variable sequence, and bring it into equations (14) to (17) to calculate the indicators of $E_{IR}$, $E_{IO}$ and $E_{IO}$.

The flow chart of flexibility margin analysis is shown in Figure 4.

# Research on Flexible Margin of Electric-Hydrogen Coupling Energy Block Based on MPC

## 5 Example simulation

### 5.1 Energy block parameter setting

The example system structure is shown in Figure 1. In the electric hydrogen coupling energy block constructed in this paper, the installed capacity of wind farm is 60MW, the installed capacity of photovoltaic power generation system is 60MW, and the installed capacity of gas turbine assembly is 30MW. In terms of hydrogen energy system, including 30MW fuel cell power generation system and 30MW electrolytic cell system, the initial SOC of hydrogen energy storage is taken as 40%. The energy storage device adopts a battery with a total capacity of 120mwh, its rated charge discharge power is ±10MW, and the initial SOC value of the battery is 45%. Refer to literature [25-28] for the parameters of matrix $B$ in equation (25).

Relevant technical parameters of electro hydrogen coupling system are shown in Table 1.
The data of wind power, photovoltaic, power load and hydrogen load are shown in Figure 5.

### 5.2 Flexibility margin analysis of energy block

In order to analyze the differences of flexibility margins in different scenarios, this section designs three calculation scenarios to evaluate the flexibility margins of fast energy.

Scenario 1: In the absence of controllable components, analyze the lack of flexibility margin in the initial operating state.

Scenario 2: Add hydrogen storage and analyze the flexibility margin of the system.

Scenario 3: Based on scenario 2, battery energy storage and gas generators are added to further increase the adjustment capability of the energy block and analyze the flexibility margin of the system.

The energy block model is established according to each scene, and the system output power at the step size of 5 min is calculated and analyzed. According to the operation of each scene at each time, draw the operation diagram of energy block and the boundary envelope of dynamic flexibility margin. The flexibility margin analysis required for scenario 1 is shown in Figure 8; The simulation analysis of scenario 2 and scenario 3 is shown in Figure 6 and Figure 7, and the corresponding flexibility margin boundary envelope analysis is shown in Figure 9 and figure 10. The flexibility margin index and flexibility maximum value of each scenario are shown in Table 2 and Table 3.

In scenario 1, high permeability renewable energy is accessed, and there is no controllable element in the system to adjust. At this time, the air and light rejection can only be adjusted through the overflow energy W, resulting in the mismatch of the flexibility margin of the system and the increase of the flexibility margin index. In scenes 2 and 3, controllable elements such as hydrogen energy storage and battery energy storage are added. At this time, the flexibility margin of the energy module is effectively increased through the integration of resources. In scenario 2, the flexibility margin requirements of the module in terms of energy storage and output are supplemented through the combination of fuel cell and electrolytic cell. The operation requirement is to avoid frequent startup and shutdown of the unit on the basis of full supply of load and consumption of renewable energy. In scheme 3, the battery energy storage and gas unit further supplement the flexibility margin requirements in climbing power and unit output required by the module, further improving the flexibility margin level of the system. Scenario 2 and scenario 3 supplement corresponding units based on flexibility margin index. It can be seen from Table II and table III that through the comparison of various scenarios, the integration of controllable resources effectively increases the flexibility margin of the energy module.

As shown in Figures 9 and 10, due to the high proportion of renewable energy connected to the energy module, its strong volatility leads to insufficient up regulation or down regulation capacity at many operating points. For scenario 2, in the 5-33hours interval, there is no wind at night, the fan and photovoltaic output are insufficient, and the climbing power and unit output in the energy module are insufficient, which cannot provide sufficient flexibility resources. During this period, the maximum climbing power shortage is 3.6MW/min, the maximum output power shortage is 18.5MW, and the Maximum supply power shortage is 4.27MWh. In the 163-205hour interval, the output of fan and photovoltaic decreases sharply, the SOC of hydrogen energy storage is maintained at a low level, and the fuel cell cannot provide enough electric energy, so the normal operation of the system can only be maintained through load shedding.

The maximum value of climbing power shortage during this period is 4.67MW/min, the maximum value of output power shortage is 20.7MW, and the maximum value of power supply shortage is 36.27MWh. In the 42-46, 115, 155-165, 230, 300-310, 325-355 hours interval, energy modules cannot fully absorb renewable energy. These parts of the electricity discarded account for 7% of the total renewable energy power generation. The flexibility margin index of scenario 2 is $E_{IR}$=5.631MW/min, $E_{IO}$=26.173MW, $E_{IC}$=32.089MWh. For Scenario 3, on the basis of Scenario 2, the combination of battery energy storage and gas generator sets makes up for the corresponding flexibility margin deficiency.

It can be seen from Figure 7 that the battery energy storage has the ability to respond quickly. The rapid discharge supplements the system's demand in the climbing power dimension, while the gas-fired unit supplements the lack of output power. Scenario 3 can basically meet the upward flexibility requirements of energy modules, but in the 44, 114-115, and 155-160hours interval, there are still situations where renewable energy power generation cannot be fully absorbed. The power discarded in these parts accounts for 1.5% of the total renewable energy power generation. The flexibility margin index of scenario 3 is $E_{IR}$=0MW/min, $E_{IO}$=1.1MW, $E_{IC}$=3.62MWh. To avoid the abandonment and load shedding of renewable energy sources, we can invest in flexibility resources with corresponding adjustment capabilities and adjustment directions based on the calculation results.

## 5.3 The influence of renewable energy penetration rate on the flexibility margin of energy block

Compared Renewable energy has the characteristics of intermittent and volatility, and a high proportion of renewable energy access will have a certain impact on the flexibility of the system. According to scenario 3, study the impact of different access ratios of renewable energy on the flexibility margin, gradually increase the ratio of wind power and photovoltaic access from 0% to 50%, and calculate the abandonment rate of renewable energy by formula (46). The flexibility margin of the system under different penetration rates is shown in Table 4.

$$E_w = \sum_{t=1}^{N_T} \sum_{i=1}^{2} \frac{P_{gen,i,t}}{P_{gen,i,t} + w_{i,t}} \quad (46)$$

With the continuous improvement of renewable energy penetration, the flexibility margin of energy module gradually decreases. As can be seen from Figure 11, there is a threshold for renewable energy penetration. When the access ratio exceeds 40%, the flexibility margin index of each dimension of the energy module shows a rapid growth trend. The comparison between scenario 2 and scenario 3 shows that PV requires higher flexibility margin than wind power. However, due to the intermittent characteristics of wind power generation, the demand for flexibility margin in the dimension of climbing power is high. The comparison of different scenarios shows that when the renewable energy penetration rate exceeds 40%, the $E_{IC}$ and $E_{IO}$ indicators increase significantly. When the permeability exceeds 60%, the $E_{IC}$ index increases significantly. It can be seen that the deficiency of unit output and power supply index has a greater impact on the system flexibility margin.

When analyzing system flexibility margin, analysis indicators of different dimensions are required. According to the analysis results, increase the corresponding flexibility margin resources in each

Research on Flexible Margin of Electric-Hydrogen Coupling Energy Block Based on MPC14

dimension to ensure the economy of the system while maintaining the balance of system flexibility margin as far as possible.

## 6    Conclusion

By analyzing the power exchange characteristics of heterogeneous energy sources, this paper establishes the homogenization model of various flexibility margin resources. On this basis, the energy block model of electro hydrogen coupling is established. The flexibility margin evaluation index is proposed from the dimension of system operation, and the flexibility margin of the module is analyzed from the three dimensions of climbing power, unit output and power supply. The model predictive control algorithm is introduced to solve the problem, and then the system flexibility margin is quantitatively analyzed. The conclusions are as follows:

1）By analyzing the power exchange characteristics of heterogeneous energy, a homogenization model is established based on the energy balance relationship of each unit of the energy module. The complex model of multiple physical fields is simplified and equivalent, which lays a model foundation for the on-line calculation of flexibility margin.

2）In this paper, an optimal calculation method of energy block power balance based on MPC is proposed, and the flexibility margin of electro hydrogen coupling energy block is analyzed. In this paper, the energy block balance criterion and flexibility margin index of electro hydrogen coupling are proposed from the three dimensions of climbing, power and energy. Through the analysis of the flexibility margin of the operation point, the deficiency of the flexibility margin of the system is judged, the flexibility margin analysis of the energy module is better realized, and suggestions are provided for the system scheduling.

3）The simulation results show that the flexibility index of each operation point is different due to the different intermittent and fluctuation distribution of the energy module at each time. At the same time, nodes and moments with insufficient flexibility margin can be found according to the drawing of flexibility margin envelope of operating points. The indicators proposed in this paper can help the system coordinate the configuration and operation of flexibility resources.

4）Through the analysis of permeability, it can be seen that there is a threshold of renewable energy permeability. When the threshold is exceeded, the flexibility margin index of each dimension of the energy module shows an increasing trend. Compared with wind power, PV has higher requirements on flexibility margin.

This paper only models from the steady-state dimension, and does not involve the transient model of the system and the capacity matching of the wind solar hydrogen system. In the follow-up research, many technical and economic factors such as hydrogen storage and hydrogen sales will be mainly considered. Besides, it is also interesting to investigate the planning and energy management of microgids and integrated energy system incorporating the presented electric-hydrogen coupling energy block [29, 30].

## 7    Conflict of Interest

Author Zijiao Han is employed by Institute of Electrical Engineering, Shenyang University of Technology and Liaoning Electric Power Company, State Grid Corporation of China. Author Shun Yuan is employed by Institute of Electrical Engineering, Shenyang University of Technology and National Energy Administration. Author Yannan Dong is employed by Institute of Electrical Engineering, Shenyang University of Technology. Author Shaohua Ma is employed by Institute of Electrical Engineering, Shenyang University of Technology. Author Yudong Bian is employed by Jiangmen Power Supply Bureau of Guangdong Power Grid Corporation. Author Xinyu Mao is employed by School of Electrical Engineering, Northeast Electric Power University. All authors declare no other competing interests.

## 8 Author Contributions

Zijiao Han is mainly responsible for technical aspects, conception and design of manuscripts. Shun Yuan investigated the homogenization modeling and calculated the main methods. Yannan Dong contributed to the flexibility margin analysis. Shaohua Ma provides valuable help for calculation. Yudong Bian and Xinyu Mao participated in the production of the paper chart and the preparation of some contents.

## 9 Funding

This study received funding from "Liaoning BaiQianWan Talents Program" and Liaoning Electric Power Co., Ltd. science and technology project (2021YF-81) of State Grid. The funder was not involved in the study design, collection, analysis, interpretation of data, the writing of this article or the decision to submit it for publication. All authors declare no other competing interests.

## 10 Acknowledgments

Thanks for The Project is sponsored by "Liaoning BaiQianWan Talents Program" and Liaoning Electric Power Co., Ltd. science and technology project (2021YF-81) of State Grid, which funded us, and also to the experts and scholars who have helped us in the process of scientific research.

## 12 Supplementary Material

### 12.1 Supplementary Figures

**Figure 1** Electro-hydrogen coupling energy module
**Figure 2** Heterogeneous energy homogenization model
**Figure 3** Flexibility margin analysis
**Figure 4** Flowchart of the flexibility margin calculation method
**Figure 5** The value of load, photovoltaic plant output and wind farms output.
**Figure 6** The Simulation Analysis of scenario 2.
**Figure 7** The Simulation Analysis of scenario 3.
**Figure 8** Flexibility margin analysis of scenario 1.
**Figure 9** Flexibility margin analysis of scenario 2.
**Figure 10** Flexibility margin analysis of scenario 3.
**Figure 11** Renewable energy abandonment rate.

### 12.2 Supplementary Tables

**Table.1** Parameters of the system
**Table.2** Flexibility indicators in different scenarios
**Table.3** Maximum/Minimum Flexibility indicators in different scenarios
**Table.4** System flexibility under different permeability